# Digital interferometric demodulation of Placido mires applied to corneal topography

**M. Servin***

*Centro de Investigaciones en Optica A. C., Loma del Bosque 115, 37150, Leon Guanajuato, Mexico.*
*\*mservin@cio.mx*

**Abstract:** This paper presents a novel digital interferometric method to demodulate Placido fringe patterns. This is a synchronous method which uses a computer-stored conic-wavefront as demodulating reference. Here we focus on the experimental aspects to phase-demodulate Placido mires applied to corneal topography. This synchronous method is applied to two topographic Placido images and their de-modulated corneal-slope deformation is estimated. This conic-interferometric method is highly robust against typical "noise" signals in Placido topography such as: reflected eyelashes and iris structures. That is because the eyelashes and the iris structure are high frequency "noisy" signals corrupting the reflected Placido mire, so they are filtered-out by this method. Digital synchronous interferometry is here applied for the first time to demodulate corneal topographic concentric-rings images (Patent pending at the USPTO).

## 1. Introduction

Periodic concentric-rings Placido mires are used to measure the topography irregularities of the human cornea since 1880 [1]. Modern corneal topographers still use Placido targets, or some variations of it, for example: the ATLAS 9000, the Tomey TMS, the Astra Max, the Magellan Mapper, the Keratron, the Topolyzer, the EyeSys 3000 and the EyeSys Vista [1,2]. The Placido target is placed in front of a human eye and its reflected image is digitized with a camera placed at a center hole of the Placido's mire. The reflected Placido image is phase-modulated by the slope of the corneal irregularities. The geometry of the corneal topographer determines the radial-slope sensitivity of the apparatus [7,8,9]. Then this Placido image is *phase de-modulated* to estimate the radial slope of the corneal irregularities (typically) with respect to its closest sphere [1-9]. Papers focusing on Placido phase-demodulation are uncommon [7,8,9]; nevertheless some are disclosed in Patents [3-6]. Standard Placido demodulation locates the fringe edges along meridians of the Placido image using intensity-image processing techniques [3-9]. A set of local meridian distances between the Placido rings and a *real-valued un-modulated Placido reference* are estimated [3-9]. To increase the number of estimated slope-points, more complicated Placido targets have been proposed [2-9]. Some of these newer Placido patterns are more sensitive to the angular corneal deformations [2-9]. Finally this sparse set of estimated slope-points is integrated along meridian lines to obtain the topography deformations of the testing cornea [2-9].

In contrast synchronous interferometric methods provide holographic phase estimation at every pixel of the fringe pattern domain in a robust and straightforward way [9-11,13]. Different kinds of fringe patterns need different (complex-valued) wavefront references for their synchronous demodulation. Linear demodulation uses a plane wavefront as reference [10,11], while pixelated interferometry uses a more sophisticated reference wavefront [12]. This paper presents a new digital interferometric method for Placido images which uses a conic-wavefront as reference. As far as I know, this is the first time that conic-wavefront interferometry is used to demodulate Placido concentric-rings images. One of the most significant advantages of this conic-interferometric method with respect to previous demodulation approaches based on ring-edge detection is its immunity to eyelashes reflected by the cornea as well as "noisy" iris structures. This synchronous method filter-out most eyelashes and iris structures because they may be regarded as high-frequency noisy signals and are somewhat orthogonal to the concentric-rings Placido fringes.

In this paper no attempt was made to point out any advantage or drawback of Placido topographers in clinical applications. Neither this paper pretends to compare Placido-based corneal topographers against: Slit-Scanning Topography, Scheimplug Imaging, Ultrasound Digital Topography, or Optical Coherence Topography [1,2]. The sole purpose of this paper is to communicate an accurate and efficient interferometric phase-demodulation method for corneal Placido images. This digital synchronous method may be implemented by convolution filters in the image space, or in the Fourier domain. Placido targets may also be used to test optical wavefronts as discussed in [14].

## 2. Interferometric demodulation of Placido images using a conic-wavefront reference

The basic theory for synchronous demodulation of Placido images published in [14] is reviewed here for the reader's convenience.

Although most concentric rings Placido reference patterns have a binary profile, we will model it as a continuous cosine function,

$$\text{Placido}(\rho_P) = 1 + \cos\left(\frac{2\pi}{P_p}\rho_p\right). \tag{1}$$

Where $(\rho_P, \theta_P)$ are the polar coordinates of the Placido disk. The constant $(2\pi / P_P)$ is the radial spatial frequency of the mire at the Placido source plate. A usual model for the human cornea is composed by a sphere plus a deformation function $g(\rho_c, \theta_c)$ [8]:

$$\text{Corneal\_Shape\_}(\rho_c, \theta_c) = \sqrt{R^2 - \rho_c^2} + g(\rho_c, \theta_c). \tag{2}$$

Being $R$ the radius of the ideal cornea and $(\rho_c, \theta_c)$ its coordinates [8]. Then one may model the Placido image (at the camera plane $(\rho, \theta)$) obtained by reflection of the Placido disk over the human cornea by,

$$I(\rho, \theta) = a(\rho, \theta) + b(\rho, \theta) \cos\left[\frac{2\pi}{P}\rho + s\frac{\partial g(\rho, \theta)}{\partial \rho}\right]. \tag{3}$$

Where $s$ is the topographer's sensitivity [7,8,9]. The background of the Placido image is $a(\rho, \theta)$, and its contrast $b(\rho, \theta)$. Finally $\partial g(\rho, \theta)/\partial \rho$ is the corneal-slope deformation.

To demodulate the Placido image in Eq. (3) by digital interferometry we must first multiply it by the (complex-valued) reference wavefront $\exp[-i(2\pi \rho / P)]$, *having the same spatial frequency and center as the Placido image in Eq. (3).*

$$I(\rho, \theta)\exp\left(-i\frac{2\pi}{P}\right) = \left\{a(\rho, \theta) + b(\rho, \theta)\cos\left[\frac{2\pi}{P}\rho + s\frac{\partial g(\rho, \theta)}{\partial \rho}\right]\right\}\exp\left(-i\frac{2\pi}{P}\rho\right). \tag{4}$$

The cosine signal may be decomposed into its two complex exponentials obtaining,

$$I\exp\left(-i\frac{2\pi}{P}\right) = \left\{a + \frac{b}{2}\exp\left[i\left(\frac{2\pi}{P}\rho + s\frac{\partial g}{\partial \rho}\right)\right] + \frac{b}{2}\exp\left[-i\left(\frac{2\pi}{P}\rho + s\frac{\partial g}{\partial \rho}\right)\right]\right\}\exp\left(-i\frac{2\pi}{P}\rho\right) \tag{5}$$

Where the polar axis $(\rho, \theta)$ were omitted. Finally this product may be re-written as

$$I\exp\left(-i\frac{2\pi}{P}\rho\right) = a\exp\left(-i\frac{2\pi}{P}\rho\right) + \frac{b}{2}\exp\left[-i\left(2\frac{2\pi}{P}\rho + s\frac{\partial g}{\partial \rho}\right)\right] + \frac{b}{2}\exp\left(is\frac{\partial g}{\partial \rho}\right) \tag{6}$$

This result has three terms: the first term is a conic-wavefront $a(\rho, \theta)\exp[-i(2\pi / P)\rho]$; the second term is a wavefront having twice the reference frequency $2(2\pi / P)\rho$; the third term is the desired analytical-signal $(b/2)\exp[i\,s(\partial g / \partial \rho)]$. Finally we must filter-out the two high-frequency conic wavefronts in Eq. (6) using a low-pass filter LPF[•] to obtain,

$$\text{LPF}\left\{I(\rho, \theta)\exp\left(-i\frac{2\pi}{P}\right)\right\} = \frac{b(\rho, \theta)}{2}\exp\left[is\frac{\partial g(\rho, \theta)}{\partial \rho}\right]. \tag{7}$$

This equation shows the proposed digital interferometry method. Whenever the three spectra in Eq. (6) are well separated we obtain, error-free, the desired baseband analytical signal (Eq. (7)). The condition for spectral separability, and therefore for error-free demodulation, is the following supremum (least upper bound) for the testing radial-slope $s[\partial g(\rho, \theta)/\partial \rho]$,

$$\sup\left[\frac{\partial}{\partial \rho}\left(s\frac{\partial g(\rho, \theta)}{\partial \rho}\right)\right] < \frac{2\pi}{P}. \tag{8}$$

The final step to obtain the desired corneal slope deformation $s\partial g(\rho, \theta)/\partial \rho$ from Eq. (7) is,

$$s\frac{\partial g(\rho,\theta)}{\partial \rho} = \tan^{-1}\left[\frac{\text{Im}\{\text{LPF}[I(\rho,\theta)\exp(-i2\pi\rho/P)]\}}{\text{Re}\{\text{LPF}[I(\rho,\theta)\exp(-i2\pi\rho/P)]\}}\right]. \qquad (9)$$

Where Im[·] and Re[·] take the real and imaginary parts of their argument. Eqs. (7,9) give the proposed phase-demodulation method. The radial-slope deformation is estimated at every pixel of the measuring cornea, not just at sparse points [3-9]. The phase in Eq. (9) is wrapped modulo $2\pi$, so before radial integration to obtain $g(\rho,\theta)$, we must unwrap it.

All digital synchronous interferometric methods need a complex-valued reference as local oscillator [10-12]. In linear interferometry one uses a plane wavefront *i.e.* $\exp[i(2\pi/P)(x+y)]$ as reference [8,11]. In pixelated interferometry the reference wavefront is composed by 2x2 wavefront unit-cells tiled over the 2D space [12] as,

$$\text{Pixelated\_Reference\_}(x,y) = \left[\sum_{n\in\mathbb{Z}}\sum_{m\in\mathbb{Z}}\delta(x-2n, y-2m)\right] **\exp\left[i\begin{pmatrix}0 & \pi/2 \\ 3\pi/2 & \pi\end{pmatrix}\right] \qquad (10)$$

Where ** is the two-dimensional convolution. Finally this paper uses the conic-wavefront $\exp[i(2\pi/P)\rho]$ as reference. These wavefront references demodulate different kind of fringe patterns which: a) looks different, b) have different spectra, and c) have different phase-demodulating properties. Trying to demodulate a Placido disk using a plane wavefront [10,11] or a pixelated complex-valued reference (Eq. (10)) [12] would completely fail.

### 3. Phase demodulation of two corneal Placido images

Here two Placido images are synchronously phase-demodulated using a conic-wavefront.

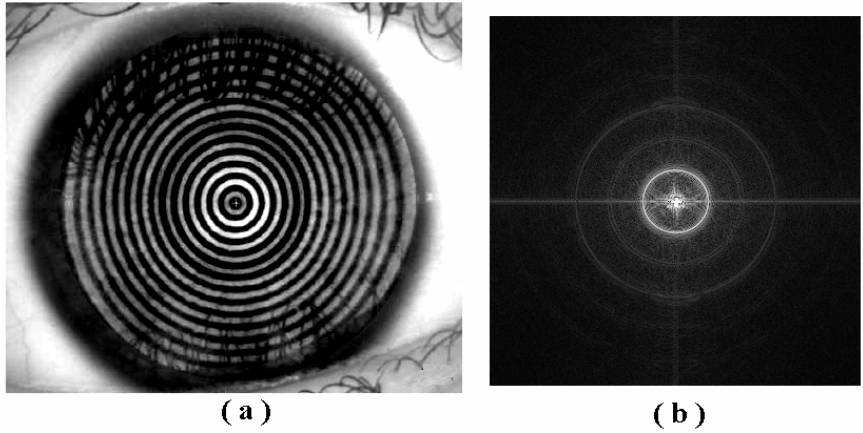

( a )   ( b )

Figure 1. Panel (a) shows the phase-modulated Placido target in Eq. (3). Panel (b) shows the spectrum of the Placido image in panel (a). Note in (b) a series of wider dim spectral rings; these correspond to higher harmonics of the binary profile of the Placido fringes.

Figure 1 shows a modulated Placido mire imaged over a human cornea and its frequency spectrum. The spatial frequency of the rings is about $(2\pi/P) = 0.6$ (radians/pixel) along the radial axis $\rho$. Panel 1(b) shows the magnitude of the Fourier spectrum of the Placido image. Please note that both, the positive and negative complex conjugate spectra are superimposed at the innermost bright-ring region in panel 1(b). In other words, the two conjugate signals are not spectrally separated as in linear interferometry [10,11], or as in pixelated interferometry [12]; at first sight, it is not obvious how to separate these two conjugate overlapping spectra.

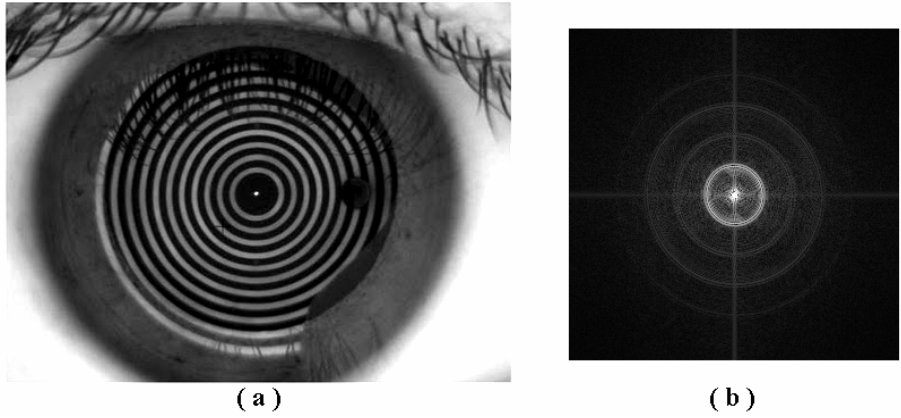

Figure 2. Panel (a) shows the phase-modulated Placido target in Eq. (3). Panel (b) shows the spectrum of the Placido image in panel (a). Note in (b) a series of wider dim spectral rings; these correspond to higher harmonics of the binary profile of the Placido fringes.

Panel 2(a) shows the corneal Placido image and panel 2(b) its Fourier spectrum. Panel 2(b) also shows the binary profile harmonics as dim higher frequency spectral rings.

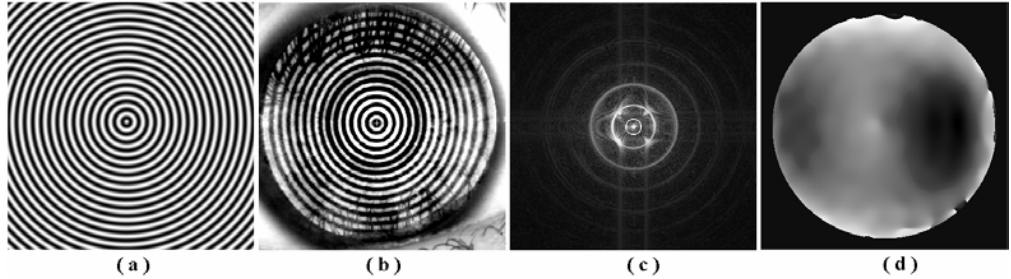

Figure 3. Panel (a) shows the real-part of the conic-wavefront used as reference to demodulate the Placido image in Panel (b). Panel (c) shows the spectrum of the Placido image multiplied by the conic wavefront; Eq. (4-6). Panel (c) also shows the size of the low-pass-filter as the bright smallest ring. Panel (d) shows the wrapped phase modulo $2\pi$ given by Eq. (8).

Panel 3(a) shows the real-part of the conic-wavefront reference and panel 3(b) its Placido image reflected over the cornea (Eq. (2)). Please note in panel 3(c) (Eq. (6)) the important fact that now: *the two conjugate spectra are well separated*. The spectrum at the origin is the desired analytical signal $(b/2)\exp[i\,s(\partial g/\partial \rho)]$. The middle-frequency spectrum corresponds to the conic-wavefront $a(\rho,\theta)\exp[-i(2\pi \rho / P)]$. Finally the outermost spectral "halo" (the middle term in Eq. (6)) is a modulated conic wavefront having a double spatial frequency $2(2\pi \rho / P)$. Panel 3(c) depicts as the innermost ring, the boundary of the $LPF[\bullet]$ used in this case. The $LPF[\bullet]$ rejects the two high-frequency conic-wavefronts in Eq. (5). Finally panel 3(d) depicts the desired phase-slope $s\,\partial g(\rho,\theta)/\partial \rho$ wrapped modulo $2\pi$ (Eq. (9)).

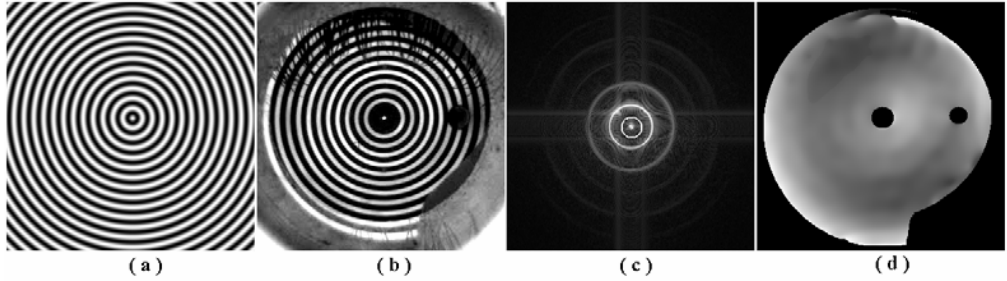

Figure 4. Panel (a) shows the real-part of the conic-wavefront used as reference to demodulate the Placido image in Panel (b). Panel (c) shows the spectrum of the Placido image multiplied by the conic wavefront Eq. (4-6). Panel (c) also shows the size of the low-pass-filter as the bright smallest ring. Panel (d) shows the wrapped phase modulo $2\pi$ given by Eq. (8).

Figure 4 has the same sequence of images but now applied to the Placido image in panel 4(b). We have masked the regions where no Placido fringes are shown, such as the one at the right where we have a dark shadow, as well as the central Placido region with no fringes. Comparing Fig. 4(b) and its demodulated phase-map in Fig. 4(d) we may see that the eyelashes where almost filtered-out by this interferometric process.

As seen from the figures herein presented, the reflected eyelashes along with other eye structures such as the iris are spurious "noisy" artifacts in corneal topography which reduces the measuring field in all Placido based topographers [1,2,7]. But as shown here they were almost filtered-out by the $LPF[\cdot]$ in this interferometric method (panels 3(d) and 4(d)). That is because the eyelashes and other iris structures are high frequency signals and are somewhat orthogonal to the Placido's rings. Also filtered-out are the high frequency harmonics of the rings profile (see Figs. 1, and Fig. 2).

Finally keep in mind that panels 3(d) and 4(d) show the deformation radial-slope $s[\partial g(\rho,\theta)/\partial \rho]$ not any corneal curvature or corneal height deformation. These may be estimated from the demodulated signal $s[\partial g(\rho,\theta)/\partial \rho]$. Also remind that: the Placido's image center and spatial frequency must equal its reference's $\exp[-i(2\pi \rho / P)]$. Otherwise an erroneous phase demodulation is obtained.

## 4. Conclusions

A new digital synchronous interferometric method for phase-demodulation of concentric-rings Placido images was presented. This method uses a conic-wavefront as reference to synchronously demodulate Placido images. This method estimates the corneal radial-slope deformation at every pixel of the Placido image's domain. This synchronous method was applied to corneal images using a Placido corneal topographer. Today intensity-based phase demodulation methods for Placido images provide only a sparse point-set of slope estimations [2-9]. Additionally intensity-image processing demodulation techniques are more complicated and prone to errors [2-9]. In contrast digital interferometry [10-12,14], provides accurate, straightforward holographic phase-demodulation at every pixel of the fringe pattern (Eq. (8)) in a single stroke. Other applications of modulated Placido patterns are: eye wavefront aberrometry, large telescope mirror testing, and adaptive optics [14]. But our main interest in this paper was to present practical aspects of Placido image demodulation in corneal topography.

A special mention must be made to the robustness of this method to demodulate Placido images containing annoying eyelashes reflected by the cornea, and other sources of "noise" such as the iris structure. This interferometric method filters-out (Eq. (7)) most eyelashes and iris structures because they are "noisy" high-frequency signals, and are almost orthogonal to the Placido rings. So the low-pass filter in Eq. (7) has no problem to eliminate them.

**Acknowledgements**

I want to thank the Mexican Science Council, CONACYT for the financial support.